# Formation and composition-dependent properties of alloys of cubic halide perovskites

Gustavo M. Dalpian[1,2*], Xingang Zhao[1], Lawrence Kazmerski[1] and Alex Zunger[1*]

[1]University of Colorado Boulder Colorado 80309, Boulder, CO, USA

[2]Centro de Ciências Naturais e Humanas, Universidade Federal do ABC, 09210-580, Santo André, SP, Brazil.

**ABSTRACT:** Distinct shortcomings of individual halide perovskites for solar applications, such as restricted range of band gaps, propensity of $ABX_3$ to decompose into $AX+BX_2$, or oxidation of $2ABX_3$ into $A_2BX_6$ have led to the need to consider alloys of individual perovskites such as $(FA,Cs)(Pb,Sn)(Br,I)_3$. This proposition creates a non-trivial material-selection problem associated with a 6-component structure, spanning a continuum of three sets of compositions (one for each sub lattice), and requiring control of phase-separation or ordering in each alloyed subfield. Not surprisingly, material and structure choices were made thus far mostly via trial-and-error explorations among a large number of arrangements. Here we use ideas from solid state theory of semiconductor alloys to analyze the behaviors of the canonical $(FA,Cs)(Pb,Sn)I_3$ alloys system, where FA is formamidinium. Density functional calculations utilizing specially constructed supercells (SQS) are used to calculate the composition dependence of band gaps, energy of decomposition and alloy mixing enthalpies. A number of clear trends are observed for A-site alloys $[Cs,FA]SnI_3$ and $[Cs,FA]PbI_3$ as well as for B-site alloys $Cs[Sn,Pb]I_3$ and $FA[Sn,Pb]I_3$. To understand the physical reasons that control these trends we decompose the alloy properties into distinct physical terms: (i) the energies associated with removing the octahedral deformations (tilting, rotations, B site displacements) of the individual components, (ii) the energies of compressing the larger component and expanding the smaller one to the alloy volume $V(x)$, (iii) the charge transfer energies associated with placing the alloyed units onto a common lattice, and finally, (iv) structural relaxation of all bonds within the cells., This analysis clarifies the origin of the observed trends in bang gaps, decomposition energies and mixing enthalpies. Unlike a number of previous calculations we find that the a proper description of alloy physics requires that even the pure, non-alloyed end-point compounds need to be allowed to develop local environment dependent octahedral deformation that lowers significantly the total energy and raises their band gaps.

## INTRODUCTION: OVERCOMING DEFICIENCIES OF INDIVIDUAL COMPOUNDS BY ALLOY FORMATION

The latest advances on the development of solar cells based on halide perovskites has shown that individual (non-alloyed) single perovskites $ABX_3$ are not optimal for this kind of application: Single $ABX_3$ perovskites have been demonstrated to have limited stability towards decomposition into binary constituents (i.e. $MAPbI_3$ transforming into $MAI+PbI_2$),[1] being prone to oxidation (i.e. $2CsSnI_3$ converting to the Sn-vacancy compound $Cs_2SnI_6$),[2–4] or loss of Halogen,[5] possible phonon (dynamic) instability,[6–8] and often not having the ideal band gaps for tandem solar applications.[9]

Because the properties $P(x)$ of an alloy can deviate significantly from the composition ($x$)-weighted linear average of the respective properties of the constituents ('alloy bowing'[10,11]), this raises the possibility that alloying might solve the problems with respect to instabilities to the above noted defficiencies. All recent record breaking solar cells[9,12–15] were made by mixtures of several perovskites instead of single compounds. This prospect of formation of alloys ('solid solutions') such as $(FA,Cs)(Pb,Sn)(Br,I)_3$, among many others,[16–25] creates a non-trivial material-selection problem associated with a 6-component structure, spanning a continuum of compositions (x, y, and z for each sub lattice), and requiring control of phase-separation or ordering in each alloyed subfield. Not surprisingly, material and structure choices were made thus far mostly via trial-and-error explorations among a large number of arrangements.[9,12–15,26] The potential for missing the best combinations or failing to apply understanding-based property optimization can be not negligible.

We will evaluate the alloy bowing for three of these properties, considering alloys based on $(FA,Cs)(Pb,Sn)I_3$. The questions on optimizing properties by creating alloys can be illustrated as follows: (i) the $APbX_3$ compound tends to decompose into $PbX_2 + AX$, whereas the $ASnX_3$ compounds do not. However, the latter have a tendency to oxidize (whereby two formula units of $ASnX_3$ form $A_2SnX_6$). Are there other ways to overcome the decomposition tendency for Pb compounds? (ii) $A[B_{1-x}B'_x]X_3$ alloy could phase-separate below a miscibility gap (MG) temperature $T_{MG} \sim \Delta H_{mix}(x)/S$ into its alloyed constituents $xABX_3 +(1-x)AB'X_3$ if the corresponding mixing enthalpy delta $\Delta H_{mix}(x)$ is positive.[27] Would halide perovskite alloys be resilient to phase separation at a sufficiently low $T_{MG}$, or would the alloy order crystallographic below an ordering temperature $T_{ord}$ (if delta H(x) is negative)?

The analysis of the bowing properties will be made through the a Born-Haber-like cycle, where the variation of the properties will be broken into a series of sequential steps, that will provide understanding in finer details of the reasons for the specific behavior of the alloy.

## THE GENERALIZED ALLOY BOWING PROBLEM

The general bowing of alloy properties, including band gaps, decomposition energy and mixing enthalpy are illustrated schematically in Figure 1. The dashed lines show the reference linear interpolation, whereas the red and the green curves indicate a few possible realizations for these actual bowing curves, each with different bowing parameters. From a quantitative point of view, each of these properties can be described as:

**(a) *Bowing of the Band gaps***, Illustrated in Figure 1a. This is the most frequently studied deviation from linearity that is observed in alloys. For example, GaAs mixed with GaN leads[28] to large non-linearities in the variation of the band gap as a function of concentration. We define the band gap bowing parameter $b_g$ for a specific alloy $A(B_xB'_{1-x})X_3$ as:

$$E_g(x) = (1-x)E_g[ABX_3] + xE_g[AB'X_3] - b_g x(1-x) \quad , \quad (1)$$

where $b_g > 0$ (positive bowing coefficient) leads to a downward concave $E_g(x)$. Recall that bowing is generally defined with respect to a given crystal structure at all compositions; different bowing curves for different crystal structures are used to judge cases where there are structural phase transitions. The cubic (high temperature) phase of halide perovskites holds a special place, as solar cells are based on this structure that tends to have solar gaps, whereas the low temperature phases (orthorhombic, tetragonal) have gaps that are often too large for solar absorption.

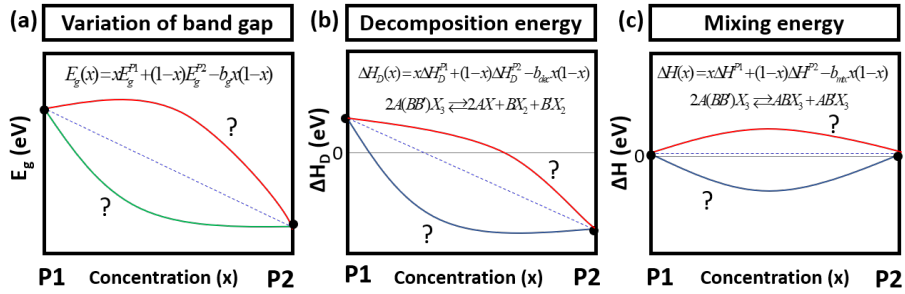

**Figure 1** Schematic description of different possible bowing situation: (linear: (blue dash line, positive (green solid line) and, negative ( red solid line) of different physical properties. (a) band gaps (b) decomposition energy ($ABX_3$ into $AX+BX_2$) and (c) bowing of alloy mixing enthalpy.

The assumed structures have a large effect on the calculated bowing. For example, the work of Goyal et al [29] treated non-cubic cell lattice vectors for the isolated compounds (as evident, for example, from the non-cubic angles between the lattice vectors), whereas their model for the cell-internal atomic positions in the alloy showed nearly ideal, weakly relaxed (non-tilted or weakly tilted) octahedra and fully non-tilted molecules. Such nearly undeformed structures produce large interaction between the states of the constituents, leading to large bowing parameters (~1.0 eV). The present work focuses attention on bowing in cubic alloys (i.e. the cell external lattice parameters are restricted to cubic, as in the physical macroscopic alloy), whereas the cell internal atomic positions are fully relaxed (to get meaningful relaxation one must initially nudge the atomic positions by random displacemets as described in the Methods section). The resulting strongly relaxed atomic positions (Table I) show in the alloy reduced interaction between the states of the individual constituents, leading to much smaller bowing parameters (~0.3 eV).

**(b) *Bowing of the decomposition energy***, related to the transformation of the ternary halide perovskite $ABX_3$ into binary compounds like AX and $BX_2$, as presented in Figure 1c. This is what has been experimentally observed in calorimetric studies.[1] The general way of evaluating the decomposition energy of an alloy is

$$\Delta H_D[A(B_xB'_{1-x})X_3] = E[A(B_xB'_{1-x})X_3] - E[AX] - (1-x)E[B'X_2] - xE[BX_2]. \quad (2)$$

Here, E[Q] is the total energy of compound Q, and the positive values of the decomposition energy mean that there is a tendency for the compound to decompose into the binaries. The respective bowing parameter ($b_{dec}$) can be obtained by fitting the curve shown in Figure 1b to the following equation:

$$\Delta H_D(x) = (1-x)\Delta H_D[AB'X_3] + x\Delta H_D[ABX_3] - b_{dec}x(1-x). \quad (3)$$

**(c) *Bowing of the alloy mixing enthalpy***: The mixing enthalpy measures the tendency of the alloy $A(B,B')X_3$ to phase segregate into its parent compounds $ABX_3$ and $AB'X_3$, as shown in Figure 1c. When this quantity is positive we expect phase separation, while when it is negative we expect ordering. This is the traditional way to account for the stability of an alloy.[30] For an alloy with mixed A-site elements $AA'BX_3$ the mixing enthalpy is calculated as

$$\Delta H[A_xA'_{1-x}BX_3] = E[A_xA'_{1-x}BX_3] - xE[ABX_3] - (1-x)E[A'BX_3]. \quad (4)$$

We can also determine a bowing parameter $b_{mix}$ for the mixing enthalpy that can be obtained through the following equation:

$$\Delta H(x) = (1-x)\Delta H[ABX_3] + x\Delta H[A'BX_3] - b_{mix}x(1-x). \quad (5)$$

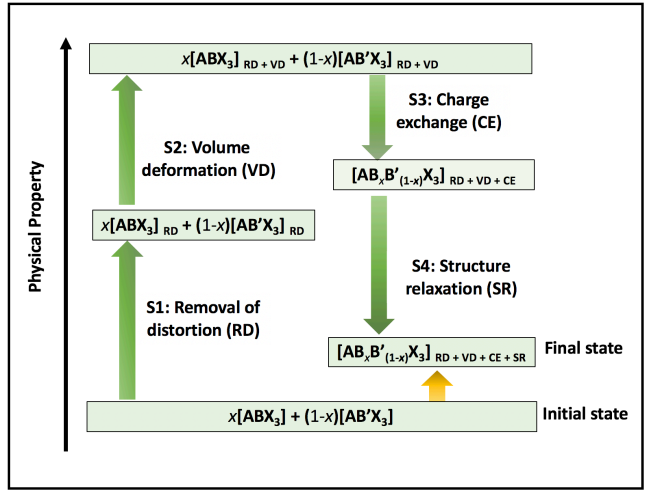

**Figure 2** Schematic diagram of the Born-Haber cycle for the formation of an alloy starting from isolated non-alloyed compounds. Illustration given for mixing B-site atoms. This cycle has two different routes: route 1, indicated by the yellow arrow that goes from the initial state directly towards the final state. In route 2, indicated by green arrows, we disentangle the alloy formation into four different steps: (S1) removal of distortions (RD), (S2) volume deformation (VD), (S3) charge exchange (CE) and (S4) structural relaxation (SR).

## DECOMPOSING THE ALLOY PROPERTIES INTO INDIVIDUAL PHYSICAL MECHANISMS

We analyze the directly calculated bowing by decomposing the energy to physically recognizable steps. Figure 2 shows illustrates the Born-Haber-like cycle we used. It starts from two compounds (illustrated by $ABX_3$ and $AB'X_3$) and define two different routes for mixing them. In the first *Direct Route* (yellow arrow) we go directly to the alloy $AB_xB'_{1-x}X_3$, comparing just the initial and final properties. In *route 2* (Born-Haber-like; green arrows) we break the reaction into four different steps, the sum of which gives the same final result as in route 1. The individual steps are:

**Step 1**: *Removal of distortions* (RD): we remove all distortions present in the individual compound (tilting, rotations, B-site off-center displacements, molecular rotations), taking all atoms into a perfect cubic perovskite structure (Pm-3m). The change in energy or the change in the physical property (band gap, decomposition energy, etc) in this step reflects the effect of undoing octahedral deformations and molecular alignment. This step is unique to perovskites, being absent from alloys of rocksalt or zinc blende compounds that do not have corner sharing octahedral.

**Step 2:** *Volume deformation* (VD): we expand the volume of smaller compound and compress that of the larger compound to fit them to the alloy volume V(x), assuming the Vegard law for the interpolation. This *Volume deformation (VD) step* can be described by the composition-weighted sum of the partial reactions

$$ABX_3|_{a_0(ABX_3)} \rightarrow ABX_3|_{a(x)} \quad (6)$$

and

$$AB'X_3|_{a_0(AB'X_3)} \rightarrow AB'X_3|_{a(x)}. \quad (7)$$

Here the subscript '0' means the equilibrium lattice constant. The change in physical property in this step reflects the effect of volume deformation of compounds that have no octahedral tilting, rotation and B-site off-center.

**Step 3:** *Charge Exchange* (CE) represents the mixing of the previously prepared equal volume components onto a fixed lattice of the alloy. The change in physical properties is described by the formal reaction

$$(1-x)ABX_3|_{a(x)} + xAB'X_3|_{a(x)} \rightarrow AB_{1-x}B'_xX|_{a(x)}. \quad (8)$$

Here we join the volume-deformed structures together to form the alloy at lattice constant $a_0(x)$. At this step the A-X and B-X bonds coexist in the alloy and all bonds are equal to each other, while charge exchange can occur among different atomic sites. CE step represents the effect of charge transfer from the less electronegative to more electronegative entity at constant volume and fixed structure.

**Step 4:** *Structure relaxation* (SR): here we allow all atoms to move into their minimal energy position with a fixed supercell volume, after giving a random displacement on the inorganic atoms, accounting for all the distortions that are present in these compounds. This *structure relaxation step* is described by the formal reaction

$$AB_{1-x}B'_xX|_{a(x)} \rightarrow AB_{1-x}B'_xX|_{a(x),rel}. \quad (9)$$

Note that for each composition the different bond lengths are not single-valued but having length distributions due to the polymorphic local environment effect, i.e., $R^{(n)}(x)_{A-X}$ and $R^{(n)}(x)_{B-X}$ are ($n$)-dependent, where $n$ refers to a given local structure. As such, each octahedra will also have different volumes and tilting angles. The molecules will also rotate in different orientations, according to the environment they are nested in.

We can now write the total change in the property P(x) of the alloy with respect to the composition-weighted average of the properties P($ABX_3$) and P($AB'X_3$) as a sum of four effects:

$$P(x) - [xP(ABX_3) + (1-x)P(AB'X_3)] =$$
$$\Delta P_{RD} + \Delta P_{VD} + \Delta P_{CE} + \Delta P_{SR} \quad (10)$$

where the left side of the equality provides the result of the direct reaction, route 1, and the right side of the equality gives its resolution into four physical components. By construction the change $P(x) - [xP(ABX_3) + (1-x)P(AB'X_3)]$ obtained via route 1 is exactly the same as in Route 2. The B-H cycle presented in Figure 2 can be used for all alloys that will be studied here and can also be used to analyze mixing effects in most of the physical properties that will be presented. In the Supplementary information we provide tables with detailed values of the properties for all these alloys inside this B-H diagrams.

## MONOMORPHOUS VS POLYMORPHOS ABX$_3$ NETWORKS

It has been known for a long time that octahedral in cubic perovskites can rotate and tilt. We have note recently that the internal energy of the cubic halide perovskite phase can be lowered relative to the nominal 1 formula unit (f.u.) description of the ideal cubic structure by using a large supercell and letting each octahedron tilt, rotate and shift its internal B atom *in a different way than other octahedral*. We refer to the ensuing configuration as polymorphous network whereby each octahedron 'sees' a different local environment. This is true even for non-alloyed ABX$_3$. Thus, we describe even the pure ABX$_3$ phases as supercells of 32 f.u./cell, while relaxing the cell internal atomic positions to minimize the internal energy, while keeping the cell external lattice vectors cubic, just like the macroscopic phase. This requires nudging the atomic position by applying initial random displacements. Naively, one would expect that minimizing the energy of a supercell with N formula units, or a single cell with one formula unit, should result in the same total energy and band gaps, as is the case for ordinary compounds such as crystalline Si or III-V's. This is not the case for many cubic halide perovskites that have dynamically unstable phonons. We find for supercells of pure ABX$_3$ that individual octahedra tilt in different amounts on different sites, creating a range of different local environments ('polymorphous network'[28,31,32]) *that significantly lowers the total energy* and increases the band gap E$_g$. In this respect, we deviate from numerous previous alloy calculations, which used for the pure phases the ideal cubic perovskite with minimum unit cell and its higher total energy. This alternative use of a single formula unit cell for the pure phases, the so-called monomorphous representation, leads to different mixing enthalpies on account of the end point compounds having artificially high energies, as discussed in the text.

**Table I: Calculated properties of single (non-alloyed) compounds including: equilibrium volume for the monomorphous (single formula unit) structure; band gap of the polymorphous phase, decomposition energy, and average distortions in the polymorphous phase, including B-site off-center distortion (Q$_B$) and absolute octahedra tilting angle (|Δθ|) with respect to perfect structure (no distortion).**

| Material | Volume (Å$^3$/f.u) | Band Gap (eV) | Decomposition Energy (meV) | Q$_B$ (Å) | |Δθ| (°) |
|---|---|---|---|---|---|
| CsSnI$_3$ | 234.6 | 0.91 | -157 | 0.10 | 9 |
| CsPbI$_3$ | 246.9 | 1.86 | -98 | 0.07 | 10 |
| FASnI$_3$ | 252.0 | 0.94 | 23 | 0.69 | 7 |
| FAPbI$_3$ | 261.7 | 1.70 | 63 | 0.14 | 6 |

## DIRECT CALCULATIONS VS. THE BORN-HABER ANALYSIS OF BOWING EFFECTS IN HALIDE PEROVSKITE ALLOYS

In this section we will present our results for the different physical properties in different alloys. We first provide the *direct results* ("route 1" in Fig 2) showing the directly calculated bowing of each property, followed by its analysis in terms of route 2: step 1 (removal of distortions); step 2 (volume deformation); step 3 (charge exchange) and in step 4 (structural relaxation). Table I collects the basic results for the individual end-point compounds, including band gaps, equilibrium volume, decomposition energies and the amplitude of the distortions (i.e., averaged B-site off-center displacement and averaged octahedral tilting angles) obtained from the polymorphous representation of these compounds

Among the possible distortions we report: B-site off center displacement, related to the tendency of the B atom (Sn or Pb) to be away from the center of the octahedra; tilting angle, related to the rotations and tilting of the octahedra compared to the perfect perovskite structure. In both cases, the magnitude of the distortion is zero for the perfect perovskites (reference).

We will discuss separately the trends in the absolute values of the band gap, decomposition energies, mixing enthalpies, and then the trends in the deviations of the absolute values from a linearly weighted average, i.e. the bowing.

**Trends in absolute values of band gaps vs composition:** In figure 3 we report the results of the direct calculation of the absolute value of the band gap in the cubic phase as a function of concentration for each of the 4 studied alloy systems. Dashed lines indicate the linear interpolation of the band gap. Full lines are a guide to the eye for our calculated band gaps. Bowing parameters were obtained by fitting the calculated values to Eq. (1).

It is interesting to note that, in agreement with experiment,[33] the band gap of CsSnI$_3$ is smaller than that of FASnI$_3$, whereas the band gap of CsPbI$_3$ is larger than that of FAPbI$_3$. These opposing trends can be understood by noting that octahedral distortions raise the band gap (REF 21), and by analyzing the relative distortions for each of these compounds, as shown in Table I. The distortions calculated for FASnI$_3$ are much larger than those of CsSnI$_3$ (mainly off-center), leading to a larger band gap in the former. On the other hand, the trend is the opposite for the Pb compounds, having larger tilting angles in CsPbI$_3$ compound relative to FAPbI$_3$ and consequently the former larger gaps.

The basic trends in the absolute values of the gaps seen in Fig 3 are:

(i) alloys on the A sublattice have negligible composition variation in the absolute values of band gap, simply because the orbitals in the A sublattice, be that Cs or molecules, are energetically far-removed from the frontier orbitals of the B and X atoms. In contrast,

(ii) Alloying on the B sublattice has a major effect on the absolute value of the gaps simply because the –atom orbitals are directly involved in bonding, and the atomic reference energies of Sn and Pb are very different.

**Trends in bowing parameters of band gaps:** Figure 3 (shaded areas) shows the deviation of the alloy band gap from the linear average, i.e., the bowing. We see that the bowing parameters are generally small positive signaling downward bowing, with the exception of [Cs,FA]PbI$_3$ that has a small negative bowing. Alloys on the A sublattice have smaller bowing than alloys on the B sublattice.

A direct comparison of these bowing parameters with experiment, however, is complicated by the fact that experimentally one generally mixes measured gaps of different crystal structures at different compositions where phase transitions from cubic to orthorhombic occurs.[33] When we add to our calculated band gaps of the orthorhombic phases (circles on the far right of Fig 3) to the rest of the calculated cubic values it is clear that the effective bowing parameter changes. We conclude that the bowing parameters of the intrinsic cubic phases are small, and larger observed anomalous band gap bowing should be mostly related to structural changes, similar to what has been proposed before for $(FAPbI_3)_x(CsSnI_3)_{1-x}$,[34] where $FAPbI_3$ is mixed with $CsSnI_3$ and a transition from cubic to orthorhombic is observed.

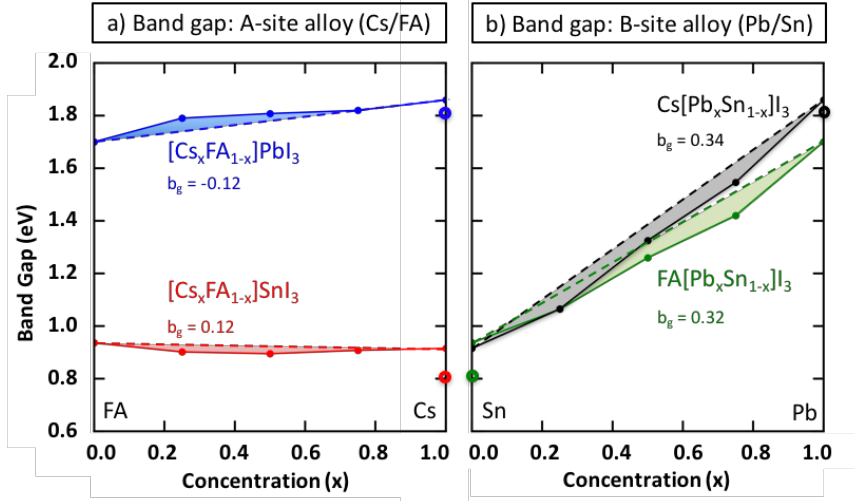

**Figure 3** Variation of the band gap as a function of alloy composition for (a) Mixed A site and (b) Mixed B site alloys. All results are calculated using a polymorphous crystal structure for both alloy and end-point compounds in the cubic phase. Dashed lines indicate the linear interpolation between the limiting compounds, and solid lines are a guide to the eye. Bowing parameter (b) obtained from parabolic fits are also given. The larger circles indicate the band gap of the orthorhombic phase of the inorganic compound

**Table II: Calculated magnitude of octahedral distortions in alloy perovskites.** Here we report the averaged off-center displacement ($Q_B$) and absolute tilting angles ($|\Delta\theta|$) with respect to the perfect structure (no distortion). We also provide the character of the valence band maximum (VBM) and conduction band minimum (CBM).

|  | $Q_B$(Å) | $|\Delta\theta|$ (°) | VBM Sn-s | VBM Pb-s | VBM I-p | CBM Sn-p | CBM Pb-p | CBM I-p |
|---|---|---|---|---|---|---|---|---|
| $Cs[Pb_{0.5}Sn_{0.5}]I_3$ | 0.11 | 9 | 0.21 | 0.06 | 0.44 | 0.30 | 0.20 | 0.05 |
| $FA[Pb_{0.5}Sn_{0.5}]I_3$ | 0.33 | 7 | 0.20 | 0.06 | 0.44 | 0.29 | 0.25 | 0.02 |
| $[Cs_{0.5}FA_{0.5}]SnI_3$ | 0.32 | 8 | 0.30 | -- | 0.43 | 0.51 | -- | 0.03 |
| $[Cs_{0.5}FA_{0.5}]PbI_3$ | 0.12 | 10 | -- | 0.20 | 0.46 | -- | 0.53 | 0.03 |

Figure 4 provides the variation $P(x)-[xP(ABX_3) + (1-x)P(AB'X_3)]$ for the band gap of the alloys in terms of the 4 contributions defined in the Born-Haber cycle ($\Delta P_{RD}+\Delta P_{VD}+\Delta P_{CE}+\Delta P_{SR}$). For each alloy, the first panel of the figure presents the total band gap change, as indicated in Route 1 whereas the change in the band gap owing to the other contributions are denoted in the following panels.

We see that:

(i) The generally small band gap bowing in halide perovskite alloys results from vanishingly small volume deformation (VD) and charge exchange (CE) contributions, and the effective compensation of removals of octahedral distortions (RD) versus the bond relaxation (SR) in the alloy. Regarding the latter compensation, the (positive) SR effect is smaller in magnitude than the (negative) RD effect, so the net change $P(x) - [xP(ABX_3) + (1-x) P(AB'X_3)]$ is negative.

(ii) Contrary to the alloys of conventional semiconductors, where the VD term is dominant[28] so that alloys of ~2-5 % lattice constant mismatch have very large bowing, in the corresponding perovskites there is a strong strain relief in terms of the distortions of the corner sharing octahedra that obviate the effects of strain. This is represented by the small VD term and the effective compensation of RD and SR.

(iii) The larger bowing in the mixed B-atom alloys (Sn and Pb) than in the mixed A site alloys (FA and Cs) results, in part, from the existence of charge exchange in the former case (open shell Sn and Pb ions as opposed to the neutral molecules).

*(iv) The role of charge separation*: Goyal et al noted that there is charge separation in the alloy band edges whereby the VBM is localized on Sn-*s* orbitals whereas the CBM is localized on both Pb-*p* and Sn-*p* (slightly more on Pb,-*p*).They argued that this charge separation explains the large (~1.0 eV) bowing parameters they calculated as the VBM inherits its character from ASnX$_3$ and the CBM inherits its character from APbX$_3$. Our calculation shows a significant charge separation (see Table II), but we do not find large bowing for the cubic alloy (see Figure 3). Hence, charge separation can not be the cause of bowing magnitude.

The Born Haber analysis of Fig 4 provides the full analysis of the bowing. We see from Fig 4 that the RD+VD+CE steps produce almost zero bowing, and it is the last step of structural relaxation (SR) that gives the final small nonzero bowing. In fact, because we use the minimum energy polymorphous representation both for the alloy and the pure compounds, we find significant octahedral distortions both for the pure compounds and the alloy, so the total bowing, being a differential quantity, is rather small. Furthermore, the charge separation seen in the alloy is not necessarily inherited from the constituents because in our calculation both are deformed.

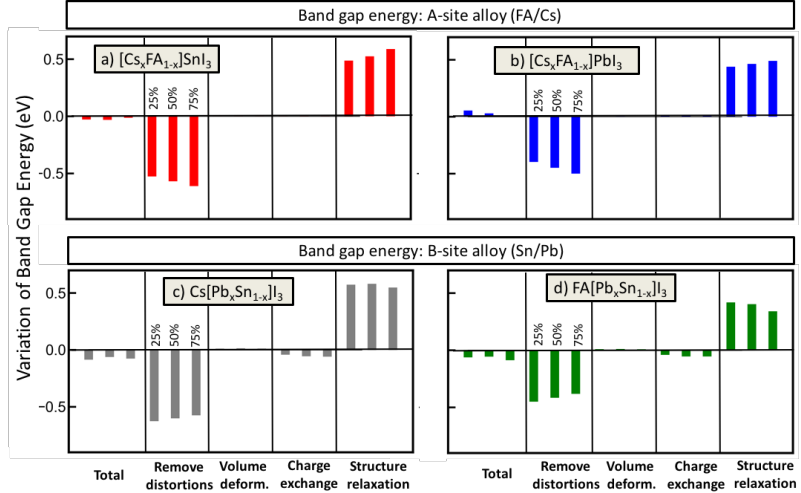

**Figure 4:** Decomposition of the total **band gap** change with respect to the composition weighted average P(x) −[xP(ABX$_3$) + (1-x) P(AB'X$_3$)] into the four terms of the Born-Haber cycle $\Delta P_{RD}+\Delta P_{VD}+\Delta P_{CE}+\Delta P_{SR}$ as described in Figure 2. All results are calculated using a polymorphous crystal structure for both alloy and end-point compounds in the cubic phase. The seemingly empty bars in VD and CE represent small values. Each effect is shown at three compositions (x=25%, 50% and 75%).

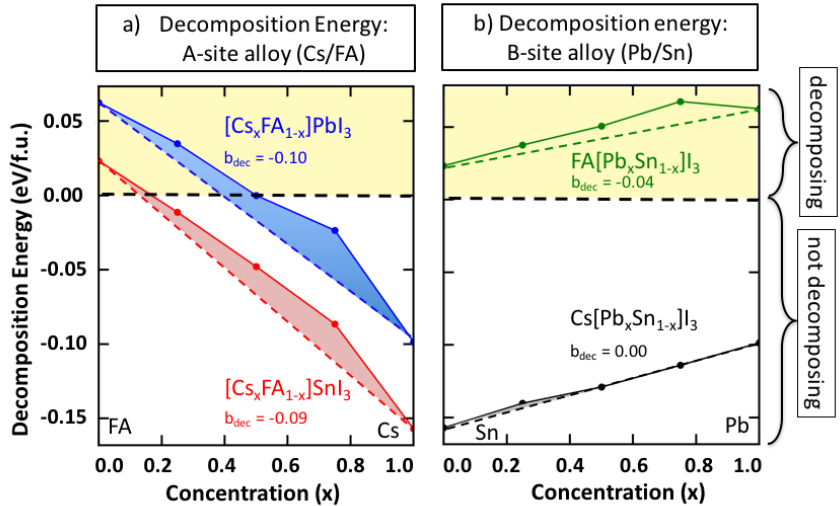

**Figure 5:** Variation of the absolute decomposition energy as a function of alloy composition for (a) Mixed A site and (b) Mixed B site alloys. All results are calculated using a polymorphous crystal structure for the cubic phase. Dashed lines indicate the linear interpolation among the limiting compounds, and solid lines are a guide to the eye. Bowing parameter obtained by fitting the results to Eq. 3. The yellow shading indicates positive decomposition energies, i.e., where the sample will break into AX+BX$_2$.

**Trends in absolute values of decomposition energies vs composition:** The decomposition into binary compounds is an important cause of instability in single ABX$_3$ halide perovskites and has been accessed experimentally by using calorimetric methods, as recently shown for MAPbI$_3$.[1] This specific material has been shown to be unstable with respect to decomposition, with a measured formation enthalpy (or decomposition energy as defined in our manuscript) of 357meV/f.u.

Table I provides the decomposition energies of the end point compounds. Positive decomposition energies signal decomposition, whereas negative values imply no decomposition. We see that FA compounds are predicted to decompose, whereas Cs compounds should not. For the same A and X atoms, Sn compounds are more resilient to decomposition than Pb compounds. This has been reported by us before,[35] and can be understood by the increased stability of the PbI$_2$ binary and the fact that the large FA molecules increases lattice parameters, weakening the B-X bonds and destabilizing these compounds. The absolute values of the decomposition energies are shown in Figure 5, together with the respective bowing coefficients. Dashed lines indicate the linear interpolation among limiting compounds. From this data it becomes clear that:

(i) Pb-based compounds are more likely to decompose than Sn-based compounds;

(ii) Cs-based compounds are less prone to decomposition than FA-based ones;

(iii) The alloying of Cs ions into the FA-based compounds makes the compound more resistant to decomposition. In the presence of Cs, all Sn- Pb alloys will not decompose.

(iv) The inclusion of small FA concentrations leads to a large increase in the tendency towards decomposition of the compounds; In the presence of FA, all Pb-Sn alloys are decomposing.

**Trends in bowing parameters of decomposition energies:** The calculated bowing coefficients of the decomposition energy are larger for A-site alloys than B-site alloys. site.

While the results clearly indicate that alloying is a good way to make a compound more stable with respect to decomposition, they also show that, for B-site alloys, this will not be sufficient to qualitatively change one material from unstable to stable. It is likely that to stabilize Pb compounds with respect to decomposition one needs to enhance the Cs alloying.

Figure 6 provides the variation $P(x)–[xP(ABX_3) + (1-x)P(AB'X_3)]$ for the alloy decomposition energies in terms of the 4 contributions ($\Delta P_{RD} + \Delta P_{VD} + \Delta P_{CE} + \Delta P_{SR}$). Here again the VD and CE contributions are negligible, whereas the (negative) SR term is smaller in magnitude that the (positive) RD term so the net effect on $P(x) –[xP(ABX_3) + (1-x) P(AB'X_3)]$ is small positive. The removal of deformations (RD) causes an increase in the total energy of the compound (less stable), so decomposition becomes a likely channel. For this step, the change in the decomposition energy is larger for compounds that have organic molecules, since this large cation must fit into the restricted space of the cuboctahedral site.

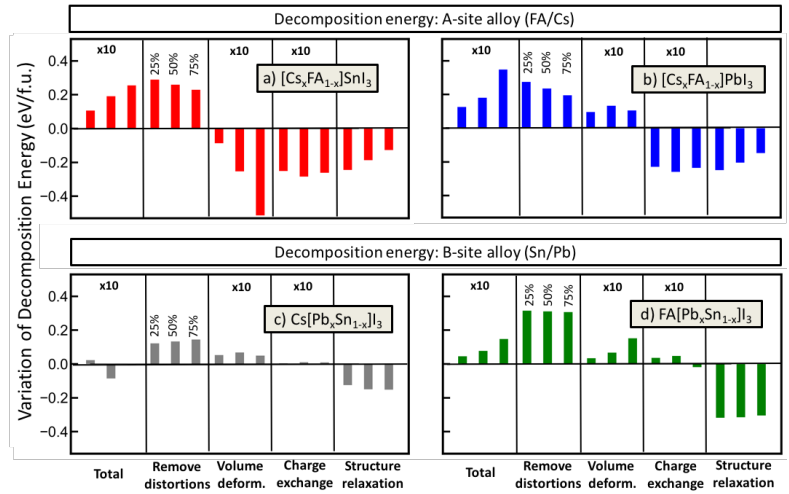

**Figure 6:** Breakdown of the *decomposition energy* of the alloy into its binary components (eg $AX+BX_2$) with respect to the composition weighted average into the four terms $\Delta P_{RD} + \Delta P_{VD} + \Delta P_{CE} + \Delta P_{SR}$ as shown in Figure 2 All results are calculated using a polymorphous crystal structure for both alloy and end-point compounds in the cubic phase Each effect is shown at three compositions (x=25%, 50% and 75%). Note that the absolute value of the bars has been multiplied by a factor of 10 in some of the panels.

**Trends in absolute values of alloy mixing enthalpy vs composition**: Special attention must be devoted to the calculation and interpretation of alloy mixing enthalpy, mainly related to the choice of the structural description of the end point compounds used to plot the energy differences in Figure 8. If one uses as a reference for the pure compounds a monomorphous representation, usually represented by a cubic, single-formula unit cell, this will lead to artificially negative values of the mixing enthalpies, concluding in favor of an ordering tendency for these systems. If instead we use a polymorphous representation, simulated by supercells where a random displacement is given to the inorganic parts of the compounds before relaxation, then we will have lower total energies to the limiting compounds, leading to positive mixing enthalpies for the alloy, predicting phase separation. This is evidenced in

Figure 7, where we plot the mixing enthalpy using two different configurations for the simulated materials. The solid curves represent energies calculated for the monomorphous structures. The calculated result is negative, indicating ordering. The dashed lines indicate the mixing enthalpy calculated using the polymorphous configurations, leading to positive mixing enthalpies. Because the polymorphous energy of the end point compound gives a lower total energy than the monomorphous, we judge the latter to the correct choice. Literature calculations show both positive and negative results for mixing enthalpies in alloy perovskites for A-site[36], B-site[29] and X-site[37] alloys. This suggests the possibility that perhaps monomorphous representation might have been used as the reference for the limiting compounds.

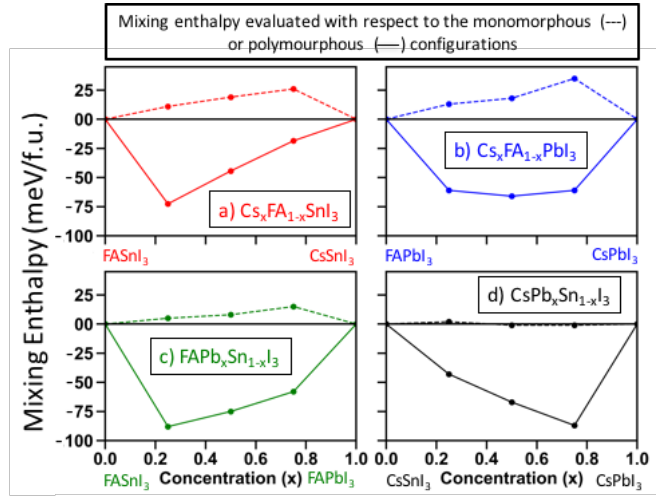

**Figure 7** Mixing enthalpy as a function of composition calculated in two ways: Using for the end-point individual perovskites a minimal cell of 1 f.u. in the ideal cubic structure (monomorphous representation) shown as solid lines, with negative mixing enthalpies implying long range ordering at low temperatures. The correct way is to use for the end point compounds a large supercell (polymorphous representation), same as used for the alloy, allowing it to relax under constraint of cubic lattice. This gives the results shown by the dashed lines—positive mixing enthalpies signaling phase separation.

As shown in Figure 8, although the positive results for the mixing enthalpy are an indication that phase separation should occur, one also can note that these values are very small, so the miscibility gap temperature will be proportionally small. Bowing parameters can be as large as 0.1eV for A-site alloys, indicating that alloying the A site induces stronger distortions on the compounds than in B-site alloys. Similar to the case of the dissociation energy, the inclusion of a small fraction of FA molecules into the inorganic compounds lead to a strong destabilization of the material. This occurs because the lattice parameters are mostly determined by the inorganic Cs atoms, leaving small spaces to fit the large FA molecules, leading to an increase in the energy of the compound.

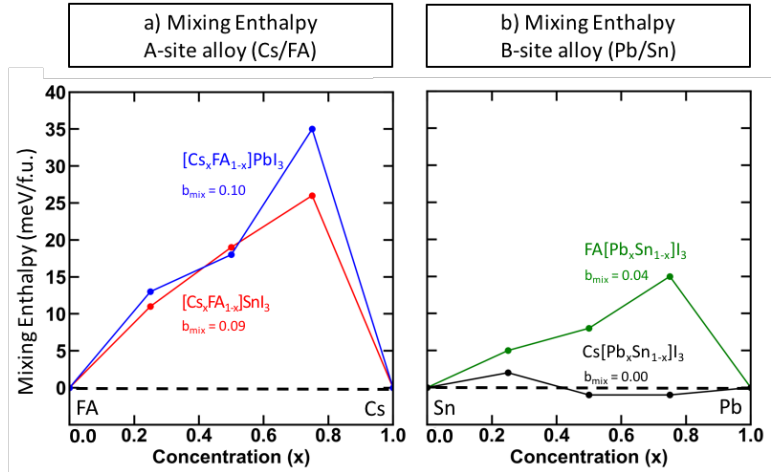

**Figure 8** Variation of the mixing enthalpy as a function of alloy composition for (a) Mixed A site and (b) Mixed B site alloys. All results are calculated using a polymorphous crystal structure for the cubic phase. Dashed lines indicate the linear interpolation among the limiting compounds, and solid lines are a guide to the eye. Bowing parameter obtained by fitting Eq.(8) are also indicated. Positive values indicate tendency for phase separation.

**Trends in the bowing of the mixing enthalpy:** Figure 9 provides the variation $P(x) - [xP(ABX_3) + (1-x) P(AB'X_3)]$ for the mixing enthalpies in terms of the 4 contributions $\Delta P_{RD} + \Delta P_{VD} + \Delta P_{CE} + \Delta P_{SR}$. The effect of removing the distortions is a large increase in the mixing enthalpy, similar to the case of the decomposition energy. Removing these distortions and consequently going to a monomorphous representation largely increases the total energy of the system.

The effect of Volume deformation will depend on the specific characteristics of each compound. As the previous step changed the potential energy surface where this material is nested, in the new potential energy surface the material might be in the minimum volume or not. If the minimum energy volume is still the same, the contribution will be positive, while if the minimum changes, the energy might decrease.

The third step of the B-H cycle is charge exchange, coming from the chemical mixture of the compounds. The change in mixing enthalpy owing to charge transfer is much larger in A-site alloys than in B-site alloys.

The last step, that is structural relaxation, is mostly a compensation of the first step: it always leads to a very large negative contribution owing to the large total energy decrease it induces in the system. For the case of (Cs,FA)SnI$_3$, the total change in mixing enthalpy is positive; the only step that has positive contribution in the decomposed cycle is RD, indicating that this is the dominant effect. The interpretation of this is that the FA molecules have more freedom to move inside the pure compounds that on alloys, restricting the energy gain in this case, and leading to positive mixing enthalpy. For the case of Cs(Pb,Sn)I$_3$, on the other hand, the bowing parameter is basically zero, because both the alloy and the pure compounds have the same ability to relax (no organic molecules), making the contribution of RD and SR basically the same. The order of the mixing enthalpies, i.e.,

[Cs,FA]PbI$_3$ > [Cs,FA]SnI$_3$ > FA[Pb,Sn]I$_3$ > Cs[Pb,Sn]I$_3$.

can be understood by the compensation among SR and RD: the difference among both is smaller for Cs[Pb,Sn]I$_3$, followed by FA[Pb,Sn]I$_3$, [Cs,FA]SnI$_3$ and [Cs,FA]PbI$_3$.

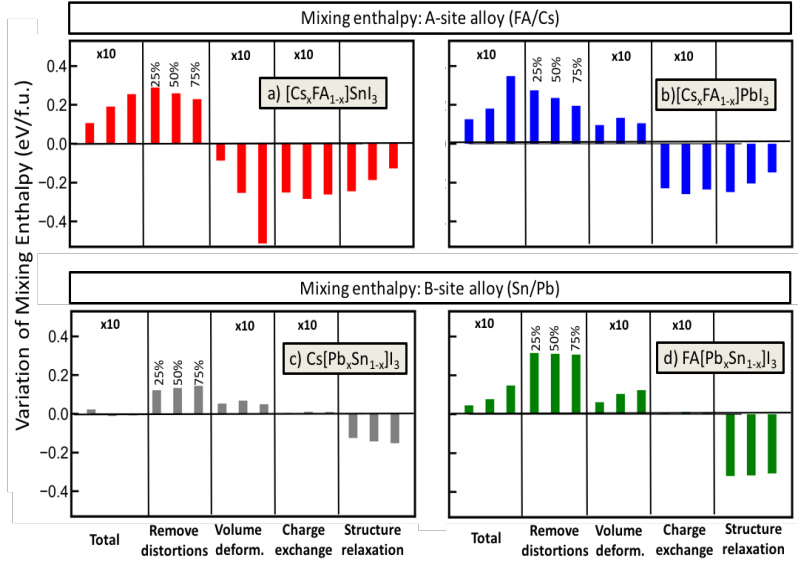

**Figure 9** Decomposition of the mixing enthalpy with respect to the composition weighted average into the four terms $\Delta P_{RD} + \Delta P_{VD} + \Delta P_{CE} + \Delta P_{SR}$, as described in Figure 2. All results are calculated using a polymorphous crystal structure for both alloy and end-point compounds in the cubic phase. Each effect is shown at there compositions (x=25%, 50% and 75%).

## CONCLUSIONS

Building alloys of perovskites compounds by mixing A, B and X elements in ABX$_3$ is a very effective way to tune the properties of these alloys, and also to overcome some drawbacks of single compounds, such as tendencies to decomposition. A given property of these alloys is usually in between the properties of the limiting compounds. However, they usually deviate from a linear interpolation among the limiting cases, showing some characteristic bowing.

We have analyzed the bowing properties of these alloys by using a Born-Haber cycle, and also using a systematic approach to consider the polymorphous nature of these materials. With this B-H cycle, we can precisely track the chemical and structural origins of the bowing of a certain material.

For *band gaps, decomposition enthalpies and mixing enthalpies* we observe that bowing does not come from volume deformation or charge exchange: it is a competition among the relaxation of the material in the pure compounds (RD) and in the alloys (SR).

We've also observed that the decomposition of Pb-based compounds can be reduced by inserting Cs into the materials with organic molecules. For the mixing enthalpy our results indicate small positive mixing enthalpies, that should be easily overcomed by temperature effects. Also, attention should be devoted to the use of a polymorphous representation for all materials, even limiting compounds, in order to get the correct description of the alloy.

## METHODS

*(i) Alloy structure*: We model the random alloys as well as the pure ABX$_3$ constituents by using supercells containing 32 formula units (160 atoms for all-inorganic compounds). The occupation of lattice sites by the alloyed atoms was generated via the Special Quasirandom Structures (SQS) method,[38] that assures the best random substitutional statistics possible with a given supercell size. The SQS were generated with the ATAT code[39] considering pairs and triplets with correlations of less than 10%. The volume of the alloys was taken as linear interpolation of the limiting compounds (Vegard's rule). The linear variation of the lattice parameter is supported by experimental evidences for alloys of these compounds.[34] The reference volume for each compound is shown in Table 1. We focus our study on the cubic phase of the perovskites, i.e. the overall shape of the cell was kept cubic whereas the magnitudes of the lattice vectors were adjusted to the volume at V(x). All cell internal coordinates are relaxed subject to this constrain. Force tolerance criteria for relaxation was 0.01eV/Å. The initial geometries before relaxation were taken as the perfect cubic perovskite structure; we then provide a small random displacement on the X and B atoms (different for each of the *N* independent octahedra), letting them relax to their minimal energy position. For the compounds where A=organic molecule, we selected two possible orientations ("up" or "down") for the dipole of each molecule in the supercell, and the distribution of the orientations was arranged in a random way so that the sum of all dipoles was zero. This is in agreement with recent studies that show that different *pair modes*, or arrangements of molecules, are possible.[40]

(ii) *DFT Details*: All the calculations reported here were performed using the density functional theory, using similar technical settings as in our preliminary paper.[35] We have used the projected augmented wave method (PAW) and the GGA/PBE exchange correlation functional without spin orbit coupling (SOC), as implemented in the VASP code.[41] Our PAW potentials had the following valence configuration: Cs (s2p6s1), I (s2p5), Sn (s2p2), Pb (s2p2), C (s2p2), H (s1) and N (s2p3). Cutoff energies were set to 520 eV, and k-point sampling varied depending on the supercell size (Gamma point calculations for 32 formula unit supercells). We have also used the optB86b[42] van der Waals corrections.

## ASSOCIATED CONTENT

### Supporting Information

The Supporting Information is available free of charge on the ACS Publications website.

Tables with the data used to build figures 4, 6 and 9 (PDF).
Structure files of all studied alloys in VASP format (zip).

## AUTHOR INFORMATION

### Corresponding Author

* e-mail: Gustavo.dalpian@ufabc.edu.br
* e-mail: alex.zunger@colorado.edu

**Notes**
Any additional relevant notes should be placed here.

## ACKNOWLEDGMENT

The work at the University of Colorado Boulder was supported by the U.S. Department of Energy, Energy Efficiency and Renewable Energy, Under the SunShot "Small Innovative Programs in Solar (SIPS)" Project Number DE-EE-0008153 1. This work used resources of the National Energy Research Scientific Computing Center, which is supported by the Office of Science of the U.S. Department of Energy under Contract No. DE-AC02-05CH11231. A portion of the research was performed using computational resources sponsored by the Department of Energy's Office of Energy Efficiency and Renewable Energy and located at the National Renewable Energy Laboratory. G.M.D. also thanks financial support from Brazilian agencies FAPESP and CNPq.